\documentstyle[aps,prl,epsfig]{revtex}
\begin{document}
\title{ Exact Calculation of Ring Diagrams and the Off-shell Effect on the Equation of State}
\author{Jinfeng Liao, Xianglei Zhu and Pengfei Zhuang \\
        Physics Department, Tsinghua University, Beijing 100084, China }
\maketitle \vspace{0.3in}
\begin{abstract}
\setlength{\baselineskip}{18pt} The partition function with ring
diagrams at finite temperature is exactly caluclated by using
contour integrals in the complex energy plane. It contains a pole
part with temperature and momentum dependent mass and a phase
shift part induced by off-shell effect in hot medium. The
thermodynamic potentials for $\phi^4$ and $\phi^3$ interactions
are calculated and compared with the quasi-particle (pole)
approximation. It is found that the off-shell effect on the
equation of state is remarkable.
\end{abstract}
\vspace{0.3in}
 \noindent ${\bf PACS: 11.10.Wx,24.10.Pa}$

\section {Introduction}
\setlength{\baselineskip}{19pt} The thermodynamic properties of a
system are fully characterized by its thermodynamic potential,
with which one can derive all the thermodynamic functions such as
pressure, energy density and entropy density from the well-known
thermodynamic relations. With finite temperate field
theory\cite{kapusta,bellac}, the partition function can be
calculated perturbatively for weakly coupling systems such as high
temperature phase of Quantum Chromodynamics (QCD). However, the
normal perturbation method can not describe the collective effect
which is now believed to play a very important role\cite{blaizot}
in understanding the quark-gluon plasma (QGP)\cite{qgp}possibly
formed in relativistic heavy ion collisions\cite{rhic}. It is
therefore necessary to perform resummation to include all
high-order contributions in the equation of state of the system in
hot and dense medium. In a quasi-particle description, the
in-medium effect is reflected in a temperature and density
dependent mass only. This effective mass is used to
explain\cite{quasi} the difference between the lattice
calculation\cite{lattice} of the equation of state of QGP and the
corresponding Stefan-Boltzmann limit: the particle mass goes up
with increasing temperature and then cancels partly the high
temperature effect. However, the in-medium effect changes not only
the particle mass but also its width. Most discussions concerning
thermal width are focused on its relation to particle decay, while
how it contributes to equation of state is still unclear. To study
the off-shell effect on the equation of state is the main goal of
this paper.

Ring diagrams are usually considered in the calculation of
partition functions\cite{kapusta,bellac} to avoid infrared
divergences and in the quark models\cite{njl} to form mesons at
RPA level. Normally the ring diagrams are calculated only for
static mode\cite{kapusta,bellac}, namely in the frequency sum only
the term with $n=0$ is considered. We will calculate the
thermodynamic potential with ring diagrams exactly and investigate
the off-shell contribution to the equation of state. We first
perform in general case the contour integration instead of the
frequency sum in the ring diagrams, and derive the thermodynamic
potential which contains the contributions from the quasi-particle
with temperature dependent mass determined by the pole equation
and from the scattering phase shift between the retarded and
advanced particle self energy due to off-shell effect. To
illustrate the quasi-particle and off-shell contributions to the
equation of state related to the study of QGP, we apply our
formulas to the popular models, the $\phi^4$ and $\phi^3$ theory.

\section {Formulas}

\begin{figure}[ht]
\hspace{+0cm} \centerline{\epsfxsize=4cm\epsffile{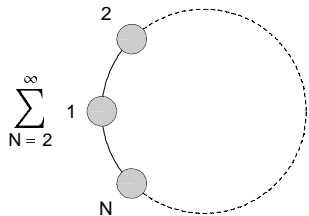}}
\caption{\it Ring diagrams with self-energy $\Pi$ indicated by
             shaded circles. } \label{fig1}
\end{figure}

The thermodynamic potential of a system with ring diagrams can be
written as
\begin{equation}
\label{otot1}
\Omega = \Omega_0 + \Omega_1 +\Omega_{ring}\ ,
\end{equation}
where \begin{equation}
\label{o0}
\Omega_0 = {1\over
\beta}\int{d^3{\bf k}\over (2\pi)^3}\ln\left(1-e^{-\beta
E_0}\right)
\end{equation}
is the free particle contribution with $E_0 = \sqrt{m^2+k^2}$
being the particle energy and $\beta=1/T$ the inverse temperature,
and $\Omega_{ring}$ is defined in
Fig.(\ref{fig1})\cite{kapusta,bellac}. To make the calculation
definite we consider in the following meson ring diagrams only.
However, the method can be straightly extended to fermions. A
shaded circle in Fig.(\ref{fig1}) means the particle self-energy
$\Pi$. To use the standard definition of $\Omega_{ring}$ we have
separated $\Omega_1$ with only one self energy on the ring from
$\Omega_{ring}$. After the summation over the rings with different
number of shaded circles, $\Omega_{ring}$ can be expressed
as\cite{kapusta,bellac}
\begin{equation}
\label{oring1}
\Omega_{ring} = {1\over 2} \int {d^3{\bf k}\over (2\pi)^3}
                {1\over \beta}\sum_n \left[\ln\left(1+A(i\omega_n,{\bf k})
                 \right)-A(i\omega_n,{\bf k})\right],
\end{equation}
where $\omega_n =2\pi n/\beta$ with $n=0,\pm 1,\pm 2,\cdots$ are
the Matsubara frequencies of meson field, and the function
$A(i\omega_n,{\bf k})$ is related to the self-energy
$\Pi(i\omega_n,{\bf k})$,
\begin{equation}
\label{aomega} A(i\omega_n, {\bf k}) = {\Pi(i\omega_n,{\bf
k})\over -(i\omega_n)^2+E_0^2}\ .
\end{equation}

\begin{figure}[ht]
\hspace{+0cm} \centerline{\epsfxsize=6cm\epsffile{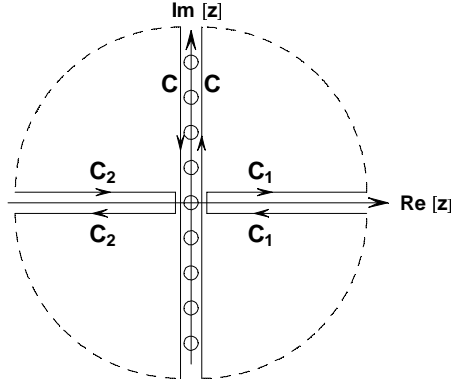}}
\caption{\it The integration contours in the complex $z$ plane. }
\label{fig2}
\end{figure}

By using analytic conjugation and the residue theorem, the
frequency summation in $\Omega_{ring}$ can be changed into an
integration along the contours $C_1$ and $C_2$ in the complex
energy plane, see Fig.(\ref{fig2}). Taking into account the
general property for the self-energy with complex energy
\begin{equation}
\label{pm} \Pi(z,{\bf k}) = \Pi(-z,{\bf k})
\end{equation}
and the asymptotic behavior
\begin{eqnarray}
\label{asy} &&  \lim_{|z|\to \infty}|z^2 A(z,{\bf
k})| < \infty\ ,\nonumber\\
&& \lim_{|z|\to \infty}|z^2 \ln \left(1+A(z,{\bf k}) \right)| <
\infty\ ,
\end{eqnarray}
the frequency summation is finally written as an integration along
the positive real axis\cite{njl},
\begin{eqnarray}
&& \label{c1} \Omega_{ring} = {1\over 4\pi i}\int {d^3 {\bf
k}\over
    (2\pi)^3} \int_0^\infty d\omega
    \left(1+{2\over e^{\beta \omega}-1}\right)\times \nonumber\\
    && \left(\ln{1+A_R(\omega,{\bf k})\over 1+A_A(\omega,{\bf k})}
    -\left(A_R(\omega,{\bf k})-A_A(\omega,{\bf k})\right)\right)\
    ,
\end{eqnarray}
where the retarded and advanced functions $A_R$ and $A_A$ are
defined by
\begin{eqnarray}
\label{ra} A_R(\omega,{\bf k}) &=& -{\Pi_R(\omega,{\bf k})
                        \over (\omega+i\eta)^2-E_0^2}\ ,\nonumber\\
A_A(\omega,{\bf k}) &=& -{\Pi_A(\omega,{\bf k})
                        \over (\omega-i\eta)^2-E_0^2}\ ,
\end{eqnarray}
with the retarded and advanced self-energies $\Pi_R(\omega,{\bf
k}) = \Pi (\omega+i\eta,{\bf k})$ and $\Pi_A(\omega,{\bf k}) = \Pi
(\omega-i\eta,{\bf k})$ and $\eta$ being an infinitesimal positive
constant .

Taking integration by parts for the logarithm term of (\ref{c1})
and using
\begin{eqnarray}
\label{araa}
&& A_R(\omega,{\bf k}) - A_A(\omega,{\bf k})
=\nonumber\\
&& {i\over 2E_0}\left( 2\pi\delta(\omega-E_0)\Pi(\omega,{\bf k}) -
{4E_0 \over \omega^2 -E_0^2}Im \Pi_R(\omega,{\bf k})\right)\ ,
\end{eqnarray}
we have (the other $\delta$ function $\delta(\omega+E_0)$ is
neglected since the integration over $\omega$ is along the
positive real axis)
\begin{eqnarray}
\label{oring3} \Omega_{ring} &=& {-1\over 4\pi i}\int{d^3{\bf
    k}\over (2\pi)^3} \int_0^\infty d\omega \left(\omega+{2\over
    \beta}\ln\left(1-e^{-\beta \omega} \right)
    \right)\nonumber\\
    &&\times{d\over d\omega}\ln{1+A_R(\omega,{\bf k})\over 1+A_A(\omega,{\bf k})}\nonumber\\
    &&- {1\over 4} \int{d^3 {\bf k}\over (2\pi)^3} \left(1+{2\over e^{\beta E_0}-1}\right)
    {\Pi(E_0,{\bf k})\over E_0}\nonumber\\
    &&+{1\over 4} \int{d^3 {\bf k}\over (2\pi)^3}
    \int_0^\infty {d\omega\over \pi }
    \left(1+{2\over e^{\beta\omega}-1}\right) \frac{2Im \Pi_R(\omega,{\bf k})}
    {E_0(\omega^2-E_0^2)}\ .\nonumber\\
\end{eqnarray}

From the separation of the logarithm,
\begin{eqnarray}
\label{sep} \ln{1+A_R(\omega,{\bf k})\over 1+A_A(\omega,{\bf k})}
&=& \ln {(\omega-i\eta)^2-E_0^2\over
(\omega+i\eta)^2-E_0^2}\nonumber\\
&+&\ln {(\omega+i\eta)^2-E_0^2-\Pi_R(\omega,{\bf k})\over
  (\omega-i\eta)^2-E_0^2-\Pi_A(\omega,{\bf k})}\ ,
\end{eqnarray}
its derivative can be expressed as a free particle part
\begin{equation}
\label{free}
{d\over d\omega}\ln {(\omega-i\eta)^2-E_0^2\over (\omega+i\eta)^2-E_0^2}
= 2i\pi\delta(\omega-E_0)\ ,
\end{equation}
and a self-energy-dependent part
\begin{eqnarray}
\label{pd} && {d\over d\omega}\ln
{(\omega+i\eta)^2-E_0^2-\Pi_R(\omega,{\bf k})\over
(\omega-i\eta)^2-E_0^2-\Pi_A(\omega,{\bf k})}\nonumber\\
&& = {d\over d\omega}\ln {\omega^2-E_0^2-\Pi(\omega,{\bf k})-i(Im
\Pi_R(\omega, {\bf k})-2\omega\eta) \over
\omega^2-E_0^2-\Pi(\omega,{\bf k})+i
(Im\Pi_R(\omega,{\bf k})-2\omega\eta)}\nonumber\\
&& = -2i{d\tilde\phi(\omega,{\bf k})\over d\omega}\
\end{eqnarray}
characterized by the phase shift $\tilde\phi$ resulted from the
difference between the retarded and advanced self-energies.

The total phase shift can be separated into two parts, a pole part
$\phi_0$ which leads to the pole equation for quasiparticle and a
scattering phase shift $\phi_s$ defined in the region $[-{\pi\over
2},{\pi\over 2}]$,
\begin{eqnarray}
\label{phi}
&& \tilde\phi = \phi_0 + \phi_s\ , \nonumber \\
&& \phi_0 = \pi\theta(\omega^2-E_0^2-\Pi)
,\nonumber\\
&& \phi_s(\omega,{\bf k}) = \arctan { Im\Pi_R(\omega,{\bf
k})-2\omega \eta \over \omega^2-E_0^2-\Pi(\omega,{\bf k})}\ ,
\end{eqnarray}
where we have made use of the property $Im\Pi_R(\omega,{\bf k})
\le 0$, which can be observed from the relation between
$Im\Pi_R(\omega,{\bf k})$ and the decay rate \cite{kapusta,bellac}
\begin{eqnarray}
\label{drate} \omega {{d \Gamma} \over {d^3 {\bf k}}} = - \frac{
Im \Pi_R(\omega,{\bf k})}{(2\pi)^3 (e^{\beta \omega}-1)} \ .
\end{eqnarray}

Substituting (\ref{free}) and (\ref{pd}) into (\ref{oring3}) and
considering the fact that the derivative of a $\theta$ function is
a $\delta$ function, we obtain
\begin{eqnarray}
\label{onon} && \Omega_{ring} = \int {d^3{\bf k}\over
(2\pi)^3}\left[
    {1\over\beta}\ln {1-e^{-\beta E} \over 1-e^{-\beta
    E_0}}
    -{1\over e^{\beta E_0}-1} {\Pi(E_0,{\bf k}) \over 2 E_0} \right]
    \nonumber\\
    && -\int{d^3 {\bf k}\over (2\pi)^3}\int_0^\infty {d\omega\over \pi}
    {1\over e^{\beta \omega}-1}\left[\phi_s(\omega,{\bf k})-
    {Im \Pi_R(\omega,{\bf k})\over \omega^2-E_0^2}\right]\
    ,\nonumber\\
\end{eqnarray}
where we have neglected the zero-point energy in the vacuum to
avoid the ultraviolet divergence. The effective energy
$E=\sqrt{m^{*2}+{\bf k}^2}$ in (\ref{onon}) is related to the
effective meson mass $m^*$ determined through the pole equation,
\begin{equation}
\label{gap} m^{*2} = m^2 + \Pi(E,{\bf k}) \ .
\end{equation}

The first term in the first square bracket of (\ref{onon}) is the
contribution from quasiparticles subtracting the corresponding
free particles which will later be cancelled in the total
thermodynamic potential $\Omega$ by $\Omega_0$, and the rest in
this bracket is an extra term resulted from the free meson
propagator between two self-energies(lines between shaded circles
in Fig.(\ref{fig1}) ). The second line of (\ref{onon}) is due to
the imaginary part of the retarded self energy which is reflected
in the scattering phase shift $\phi_s$ and an extra term resulted
also from the free meson propagators in the ring diagrams in
Fig.(\ref{fig1}). The connection between the phase of the
scattering amplitude and the grand canonical potential was first
made by Dashen\cite{dashen} and continued in \cite{norton} and
\cite{jeon}.

Now we turn to considering the lowest order correction to the
thermodynamic potential, namely $\Omega_1$ in (\ref{otot1}). Since
$\Omega_1$ differs from the second term in (\ref{oring1}) with
only a sign and a symmetry factor, it can be written as
\begin{equation}
\label{omega1} \Omega_1 = \gamma_s \int {d^3{\bf k}\over (2\pi)^3}
                {1\over \beta}\sum_n A(i\omega_n,{\bf k})
\end{equation}
where the factor $\gamma_s$ is due to the difference between the
symmetry factor for self-energy and that for partition function.
The value of $\gamma_s$ is determined by the interaction, it is $1
/ 4$ for $\phi^4$ theory and $1 / 6$ for $\phi^3$ theory with the
lowest order self-energy.

Putting together $\Omega_0, \Omega_1$ and $\Omega_{ring}$, the
total thermodynamic potential defined through (\ref{otot1}) is now
written as
\begin{equation}
\label{otot2}
 \Omega = \Omega_R + \Omega_I\ ,
\end{equation}
with the $Re\Pi$- and $Im\Pi$-dependent parts
\begin{eqnarray}
\label{oroi} \Omega_R &=& {1\over \beta}\int{d^3{\bf k}\over
(2\pi)^3} \ln (1-e^{-\beta E})\nonumber\\
&-& ({1\over 2}-\gamma_s)\int{d^3{\bf k}\over (2\pi)^3}{1\over
e^{\beta E_0}-1}{\Pi(E_0,{\bf
k})\over E_0} \ ,\nonumber\\
\Omega_I &=& -\int{d^3{\bf k}\over (2\pi)^3}\int_0^\infty
{d\omega\over \pi}{1\over e^{\beta\omega}-1}\phi_s(\omega,{\bf
k})\nonumber\\
&+& \int{d^3{\bf k}\over (2\pi)^3}\int_0^\infty {d\omega\over
\pi}{1\over e^{\beta\omega}-1}({1 \over 2
}-\gamma_s){2Im\Pi_R(\omega, {\bf k})\over \omega^2-E_0^2}\
.\nonumber\\
\end{eqnarray}
It is clear to see that $\Omega_R$ contains not only a
quasi-particle part with a temperature dependent mass $m^*$ hidden
in the particle energy $E$, but also an extra term coming from the
free meson propagator between two self-energies, and $\Omega_I$
arises from the off-shell effect which leads to a phase shift
$\phi_s$ and a similar extra term. For NJL-type\cite{njl}
interactions, there is no free propagator between two
self-energies, thus the extra terms in $\Omega_R$ and $\Omega_I$
disappear. In this case, the in-medium effect is simple and clear:
it results in a quasi-particle with a scattering phase shift. It
is also necessary to note that the quasi-particle and phase shift
are introduced only from the summation of all ring diagrams. For
any ring diagram with fixed number of self-energy $\Pi$, there are
no such contributions to $\Omega$.

\section{Examples}

With the formulas established in the last section, we can evaluate
the equation of state including resummation effect for any given
self-energy. We now consider some examples to illustrate the exact
calculation of ring diagrams and compare it with the usually used
quasi-particle approximation. The point we will focus on is the
contribution from the off-shell effect included in $\Omega_I$.

\begin{figure}[ht]
\hspace{+0cm} \centerline{\epsfxsize=6cm\epsffile{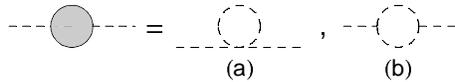}}
\caption{\it The leading order self-energies for $\phi^4 (a)$ and
$\phi^3 (b)$. } \label{fig3}
\end{figure}

Let us first consider $-\lambda\phi^4$ theory. The self-energy to
the leading order shown in Fig.(\ref{fig3}a) is evaluated
analytically for massless particles $(m = 0)$
\cite{kapusta,bellac}
\begin{equation}
\label{phi4} \Pi = \lambda T^2\ .
\end{equation}

Since the self-energy is $\omega$ and $\bf{k}$ independent, the
in-medium correction is reflected in the shift of the particle
mass only, and there is no off-shell effect. With the formulas we
develop in last section, it reads
\begin{eqnarray}
\label{phi41}
&& \Omega_I = 0\ , \nonumber\\
&& \Omega = \Omega_R = \Omega_{quasi}- {1\over 4}\int{d^3{\bf
k}\over (2\pi)^3}{\Pi\over k}{1\over
e^{k/T}-1},\nonumber\\
&& \Omega_{quasi}= \int {d^3{\bf k}\over (2\pi)^3}
T\ln (1-e^{-E/T})\ ,\nonumber\\
&& E = \sqrt{m^{*2}+k^2}\ ,\ \ \ \ \ m^{*2} = \Pi\ .
\end{eqnarray}

After integrating over the angles and scaling the momentum $k$ by
temperature $T$, the quasi-particle and total thermodynamic
potentials are both proportional to $T^4$,
\begin{eqnarray}
\label{phi42}
&& \Omega_{quasi} = a(\lambda)T^4,\nonumber\\
&& \Omega = \left(a(\lambda)-{\lambda\over 48}\right)T^4,\nonumber\\
&& a(\lambda) = {1\over 2\pi^2}\int_0^\infty dk
k^2\ln\left(1-e^{-\sqrt{\lambda+k^2}}\right)\ .
\end{eqnarray}
Note that such temperature scaling arises from the $T^2$
dependence of the effective mass $m^*$. For any other temperature
dependence, this scaling will be broken.

\begin{figure}[ht]
\hspace{+0cm} \centerline{\epsfxsize=6cm\epsffile{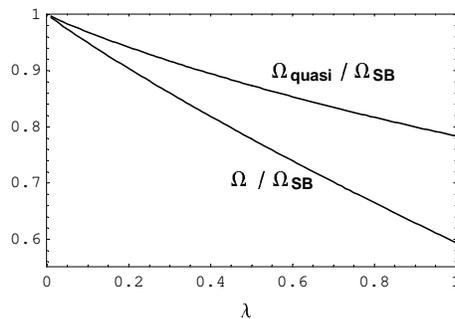}}
\caption{\it The quasi-particle and full thermodynamic potentials
scaled by the Stefan-Boltzmann limit as functions of the coupling
constant in $\phi^4$ theory. } \label{fig4}
\end{figure}

Fig.(\ref{fig4}) shows the quasi-particle and full thermodynamic
potentials scaled by the corresponding Stefan-Boltzmann limit as
functions of the coupling constant $\lambda$. We see that both the
quasi-particle and the total thermodynamics can not reach the
Stefan-Boltzmann limit, and the deviation becomes more and more
significant when the coupling constant increases. This phenomena
is fully due to the quasi-particle mass behavior: The particles
become heavy in the hot mean field and this mass transport becomes
more and more strong with increasing temperature and coupling
constant. If we take the coupling constant $\lambda$ to be about
$0.3$, the difference between the lattice calculation of the
equation of state of QGP and the corresponding Stefan-Boltzmann
limit (about 15 percent\cite{lattice}) can be accounted. We are
now certainly not handling QCD, but the spirit here and that used
in \cite{quasi} to fit successfully the lattice calculation are
quite the same.

\begin{figure}[ht]
\hspace{+0cm} \centerline{\epsfxsize=6cm\epsffile{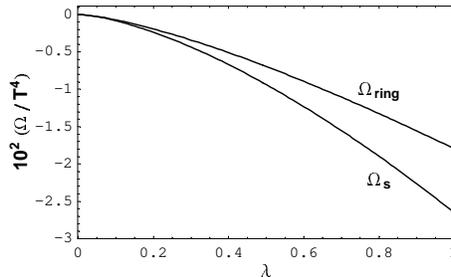}}
\caption{\it The static-mode approximation ($\Omega_s$) and the
full calculation ($\Omega_{ring}$) of the thermodynamic potential
of ring diagrams as functions of the coupling constant in $\phi^4$
theory. } \label{fig5}
\end{figure}

We introduce the so-called static-mode correction\cite{bellac},
which is obtained by taking $n=0$ only in the frequency sum in
(\ref{oring1}). The static mode plays an important role in some
recent works\cite{blaizot} as it is related to the soft degrees of
freedom in an effective theory dealing with high temperature
properties of QGP. Fig.(\ref{fig5}) shows the coupling constant
dependence of the static-mode and full calculations of the ring
diagrams. We see that only in the small $\lambda$ region the
static-mode calculation is a good approximation, for strong
interaction the contribution from the excited modes can not be
neglected.

In order to see the contribution from the off-shell effect to the
equation of state, we further consider Fock diagram in
$-\lambda\phi^3$ theory. Its self-energy to the leading order
shown in Fig.(\ref{fig3}b) can be explicitly expressed as
\begin{eqnarray}
\Pi(\omega,{\bf k}) &=& 18\lambda^2\int {d^3{\bf q}\over
(2\pi)^3}{1\over
4E_1E_2} [(1+f(E_1)+f(E_2))\nonumber\\
&& \times\left({1\over
\omega-E_1-E_2}-{1\over \omega+E_1+E_2}\right)\nonumber\\
&& -(f(E_1)-f(E_2))\nonumber\\
&& \times\left({1\over \omega-E_1+E_2}-{1\over
\omega+E_1-E_2}\right)]\ ,
\end{eqnarray}
where the factor $18$ is the symmetry factor of the $\phi^3$
theory, and $f(z)=1/(e^{\beta z}-1)$ is the boson distribution
function. The imaginary part of the retarded self-energy
\begin{eqnarray}
\label{phi31} Im\Pi_R(\omega,{\bf k}) &=& -18\pi\lambda^2\int
{d^3{\bf q}\over (2\pi)^3}{1\over 4E_1E_2}[
    (1+f(E_1)+f(E_2))\nonumber\\
&& \times\left(\delta(\omega-E_1-E_2)-\delta(\omega+E_1+E_2)
    \right)\nonumber\\
&&
-(f(E_1)-f(E_2))\nonumber\\
&& \times\left(\delta(\omega-E_1+E_2)-\delta(
    \omega+E_1-E_2)\right)]
\end{eqnarray}
can be simplified as
\begin{eqnarray}
\label{phi32} Im\Pi_R(\omega,{\bf k}) &=& -{9\lambda^2\over 8\pi
k}\int_0^\infty dq {q\over E_1}\nonumber\\
&& \times[\left(\epsilon(\omega-E_1)f(E_1)+f(|\omega-E_1|)\right)\nonumber\\
&& \times\theta(1-x_1)\theta(1+x_1)
    \nonumber\\
&& +(f(E_1)-f(\omega+E_1))\nonumber\\
&& \times\theta(1-x_2)\theta(1+x_2)]
\end{eqnarray}
with
\begin{eqnarray}
\label{phi33}
&& E_1 = \sqrt{{\bf q}^2+m^2},\ \ \ E_2 = \sqrt{({\bf k-q})^2+m^2},\nonumber\\
&& x_1 ={k^2-\omega^2+2\omega E_1\over 2kq}, \ \ \
   x_2 ={k^2-\omega^2-2\omega E_1\over 2kq}\ ,
\end{eqnarray}
and $\epsilon(x)$ being the sign function.

By substituting the self-energy and the imaginary part of the
retarded self-energy into (\ref{oroi}), we obtain a rather
complicated expression for the thermodynamic potential, which is
now impossible to get analytical result and can only be evaluated
numerically. Since the coupling constant $\lambda$ in $\phi^3$
theory is dimensional, we scale it by a momentum cutoff $\Lambda =
1 GeV$ to get a dimensionless effective coupling constant
$g={\lambda / \Lambda}$. For all the calculations below, we choose
the meson mass in the vacuum to be $m = 200 MeV$. In general, the
self-energy $\Pi$ depends separately on the variables $\omega^2$
and ${\bf k}^2$. In free space at $T=0$, it can only be a function
of the relativistically invariant combination $s=\omega^2-{\bf
k}^2$. We find, from numerical calculations, that this is also
approximately true for $T\ne 0$, and, since it introduces large
computational simplifications, we thus assume this form in the
following numerical calculations.

\begin{figure}[ht]
\hspace{+0cm} \centerline{\epsfxsize=7cm\epsffile{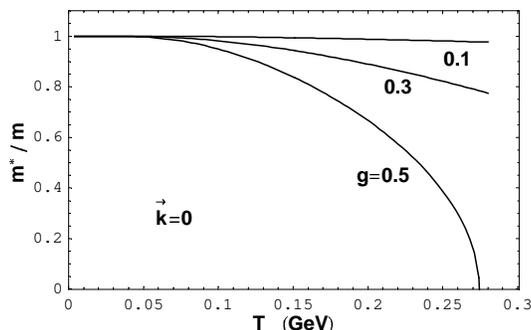}}
\caption{\it The effective mass $m^*$ scaled by the free mass $m$
as a function of temperature at ${\bf k} = 0$ for different
coupling constant $g$ in $\phi^3$ theory. } \label{fig6}
\end{figure}

\begin{figure}[ht]
\hspace{+0cm} \centerline{\epsfxsize=7cm\epsffile{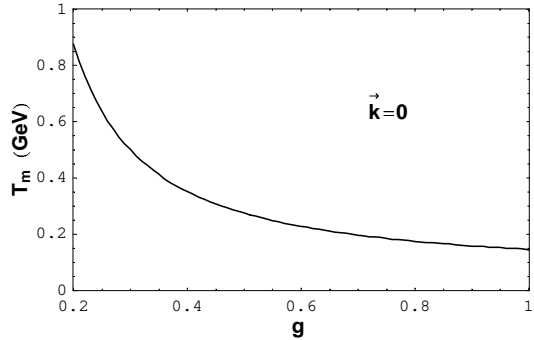}}
\caption{\it The maximum temperature $T_m$ for meson field as a
function of coupling constant $g$ at ${\bf k} = 0$ in $\phi^3$
theory. } \label{fig7}
\end{figure}

As in $-\lambda \phi^4$ theory, the summation of ring diagrams in
hot medium leads to the emergence of quasiparticles with a
temperature-dependent mass determined by the pole equation
(\ref{gap}). Fig.(\ref{fig6}) shows the temperature dependence of
the effective mass $m^*$ for different coupling constant $g$.
Unlike the $-\lambda \phi^4$ theory where the particles obtain
mass from the hot medium, the particles in $-\lambda \phi^3$
theory lose mass in the hot medium. For any given coupling
constant the effective mass drops down with increasing
temperature, and at a high-enough temperature the system finally
reaches the limit with $m^* = 0$. When the system is beyond this
maximum temperature $T_m$ determined by $m^*(T_m) = 0$, there is
no more real mass solution for the pole equation, and it leads to
an unphysical jump in thermodynamic potential (\ref{otot2}). For
the coupling constant $g=0.5$, the effective mass $m^*$ falls
quite fast and reaches zero at $T_m = 275 MeV$. We show in
Fig.(\ref{fig7}) the maximum temperature as a function of the
coupling constant $g$. We see that the maximum temperature
increases with decreasing $g$.

\begin{figure}[ht]
\hspace{+0cm} \centerline{\epsfxsize=7cm\epsffile{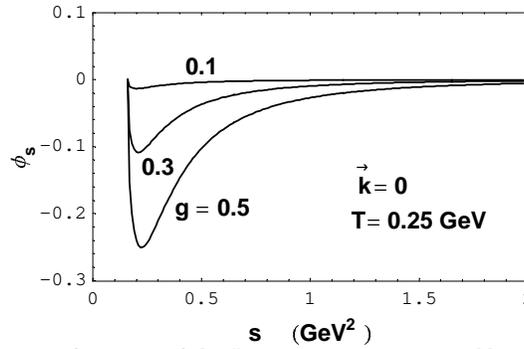}}
\caption{\it The scattering phase shift $\phi_s$ as a function of
the Lorentz-invariant variable $s$ at $T= 250 MeV$ and ${\bf k} =
0$ for different coupling constant. } \label{fig8}
\end{figure}

The resummation of ring diagrams in $-\lambda\phi^3$ theory leads
to not only an effective mass but also a scattering phase shift
due to off-shell effect. Fig.(\ref{fig8}) shows the scattering
phase shift $\phi_s$ as a function of the Lorentz-invariant
variable $s$ at $T= 250 MeV$ for different coupling constant.
Since $s-m^2-\Pi(s) < 0$ corresponding to $m^*/m < 1 $ for the
effective mass, the scattering phase shift is always in the region
$0<\phi_s < -\pi/2$ according to the definition of $\phi_s$
(\ref{phi}). It starts at the threshold energy $\sqrt s = 2 m$ for
the decay process, then reaches the maximum rapidly, and then gets
damped slowly, and finally disappears. Because the off-shell
effect in the medium arises from the thermal motion and the
interaction of particles, the scattering phase shift increases
with increasing temperature and coupling constant.


\begin{figure}[ht]
\hspace{+0cm} \centerline{\epsfxsize=6cm\epsffile{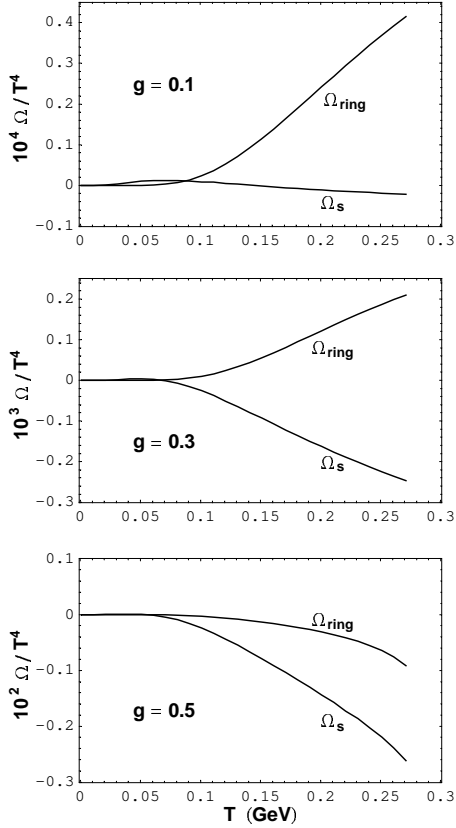}}
\caption{\it The static-mode approximation ($\Omega_s$) and the
full calculation ($\Omega_{ring}$) of the thermodynamic potential
of ring diagrams as functions of the temperature for different
coupling constant in $\phi^3$ theory.  } \label{fig9}
\end{figure}

\begin{figure}[ht]
\hspace{+0cm} \centerline{\epsfxsize=6cm\epsffile{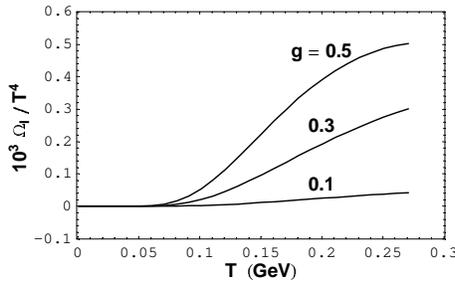}}
\caption{\it The thermodynamical potential induced by the
off-shell effect in hot medium as a function of the temperature
for different coupling constant in $\phi^3$ theory. }
\label{fig10}
\end{figure}

\begin{figure}[ht]
\hspace{+0cm} \centerline{\epsfxsize=6cm\epsffile{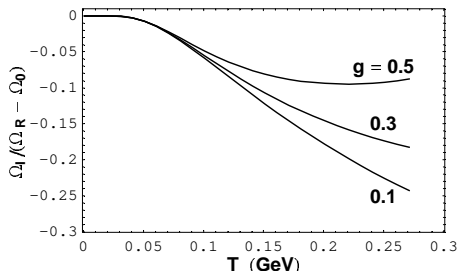}}
\caption{\it The ratio between the thermodynamical potentials
induced by the off-shell effect and by the quasi-particles in hot
medium  in $\phi^3$ theory. } \label{fig11}
\end{figure}

The off-shell effect reflected in the imaginary part of the
retarded self-energy dominates the decay process as shown in
(\ref{drate}). For thermodynamics, the thermal motion of free
particles, namely $\Omega_0$ is obviously the zeroth order
contribution, and the effective mass and scattering phase shift
result in leading order corrections. Fig.(\ref{fig9}) shows the
difference between the static-mode approximation and the full
calculation of the ring diagrams for three values of coupling
constant. While both $\Omega_s$ and $\Omega_{ring}$ increase with
increasing coupling constant (note that the scales for $g=0.1,
0.3$ and $0.5$ are different in Fig.(\ref{fig9})), the deviation
of $\Omega_{ring}$ from $\Omega_s$ is always significant, and even
more important for week couplings. This is quite different from
the case in $\phi^4$ theory, see Fig.(\ref{fig5}), where the
calculation with only static mode is a good approximation in the
region of weak coupling. This qualitative difference is mainly
from the off-shell effect in the $\phi^3$ theory. It is the
off-shell effect resulted from the frequency sum over the excited
modes that leads to the $Im\Pi$-related part in $\Omega_{ring}$,
see the second line of (\ref{onon}). Fig.(\ref{fig10}) indicates
directly the temperature and coupling constant dependence of
$\Omega_I$ induced by the off-shell effect. To see how important
the off-shell effect on the equation of state is, we plot in
Fig.(\ref{fig11}) the ratio between the thermodynamical potentials
induced by the off-shell effect and by the quasi-particles on the
mass-shell. Since the off-shell and quasi-particle effects are the
same order correction to the thermodynamical potential, we plot
the ratio $\Omega_I/(\Omega_R - \Omega_0)$ instead of
$\Omega_I/\Omega_R$. It is interesting to note that while both
$\Omega_I$ and $(\Omega_R - \Omega_0)$ get enhanced by strong
coupling, the ratio is enlarged in weak coupling case. The reason
is that the effective mass which dominates $(\Omega_R - \Omega_0)$
changes with coupling constant faster than the change in the
scattering phase shift which controls $\Omega_I$.


\section{Conclusions}

We have presented an exact calculation of ring diagrams in the
frame of finite temperature field theory in imaginary time
formalism. The resummation of the ring diagrams not only changes
the particle mass, but also generates a scattering phase shift,
both are collective effects in the hot medium. Only for the system
with self-energy in the mean field approximation, the
thermodynamics can be described by quasi-particles. In general
case, the corrections from the effective mass and from the
scattering phase shift to the equation of state are introduced at
the same time. We have applied our formulas to $\phi^4$ and
$\phi^3$ theories. For the former with the self-energy to the
first order, the static-mode can be considered as a good
approximation for weak couplings, and the enhanced mass in the hot
medium leads to the phenomena that the system can not reach the
Stefan-Boltzmann limit at high temperatures. For the latter with
the self-energy at Fock level, the suppressed mass in the hot
medium results in a maximum temperature where the particles become
massless and the system starts to collapse. The off-shell effect
in $\phi^3$ theory generates a scattering phase shift
simultaneously when the quasi-particles are formed. The total
thermodynamic potential contains a quasi-particle part and an
off-shell part. The off-shell part increases with increasing
temperature and coupling constant, while its contribution to the
total thermodynamics is enhanced in the case of weak couplings.
For the coupling constant $g = 0.1$ the ratio of the off-shell
part to the quasi-particle part reaches $25\%$ at high
temperatures. The formulas developed
here will be extended to discuss more realistic systems. \\

{\bf Acknowledgments}: This work was supported in part by the
grants NSFC19925519, 10135030 and 10105005, and the national
research program G2000077407.

\end{document}